\documentclass[authoryear,preprint,review,12pt]{elsarticle}

\usepackage{lineno}

\usepackage{hyperref} 

\usepackage[fleqn]{amsmath}
\usepackage{graphicx}
\usepackage{color}
\usepackage{caption}
\usepackage{wasysym}
\usepackage{balance}
\usepackage{longtable}
\usepackage{tabu}
\usepackage{booktabs}
\usepackage{xcolor}
\usepackage[]{units}
\usepackage{supertabular,booktabs}
\usepackage{comment}
\usepackage{amssymb}

\definecolor{hlinecolor}{RGB}{0,71,171}
\definecolor{headercolor}{gray}{0.85}

\usepackage{xcolor}
\usepackage{booktabs,colortbl, array}
\definecolor{rulecolor}{RGB}{0,71,171}
\definecolor{tableheadcolor}{gray}{0.92}
\journal{Icarus}

\begin{document}

\begin{frontmatter}



\title{Modification of icy planetesimals by early thermal evolution and collisions: Constraints for formation time and initial size of comets and small KBOs}


\author[inst1]{Gregor J. Golabek}
\ead{gregor.golabek@uni-bayreuth.de}
\author[inst2]{Martin Jutzi}
\ead{martin.jutzi@space.unibe.ch}
\address[inst1]{Bayerisches Geoinstitut, University of Bayreuth, Universit{\"a}tsstrasse 30, 95440 Bayreuth, Germany}
\address[inst2]{Physics Institute, Space Research and Planetary Sciences, University of Bern, Gesellschaftsstrasse 6, 3012 Bern, Switzerland}


\begin{abstract}
Comets and small Kuiper belt objects are considered to be among the most primitive objects in the solar system as comets like C/1995 O1 Hale-Bopp are rich in highly volatile ices like CO. It has been suggested that early in the solar system evolution the precursors of both groups, the so-called icy planetesimals, were modified by both short-lived radiogenic heating and collisional heating. Here we employ 2D finite-difference numerical models to study the internal thermal evolution of these objects, where we vary formation time, radius and rock-to-ice mass fraction. Additionally we perform 3D SPH collision models with different impact parameters, thus considering both cratering and catastrophic disruption events. Combining the results of both numerical models we estimate under which conditions highly volatile ices like CO, CO$_2$ and NH$_3$ can be retained inside present-day comets and Kuiper belt objects. Our results indicate that for present-day objects derived from the largest post-collision remnant the internal thermal evolution controls the amount of remaining highly volatile ices, while for the objects formed from unbound post-collision material the impact heating is dominant. Finally we apply our results to present-day comets and Kuiper belt objects like 67P/Churyumov-Gerasimenko, C/1995 O1 Hale-Bopp and (486958) Arrokoth.\\
\end{abstract}


\end{frontmatter}


\section{Introduction}
\label{sec:introduction}
Comets are derived from icy planetesimals and are considered to be among the most primitive leftovers of planet formation (e.g. Weissman et al., 2020) derived from either the Oort cloud (Oort, 1950) or the scattered disk population of the Edgeworth-Kuiper belt (Edgeworth, 1949; Kuiper, 1951). The study of comets like 67P/Churyumov-Gerasimenko, C/1995 O1 Hale-Bopp and 81P/Wild 2 and Kuiper belt objects (KBOs) like (486958) Arrokoth shows that these objects tend (i) to be highly porous (Sierks et al., 2015; Grundy et al., 2020), (ii) to often contain large amounts of highly volatile ices (Biver et al., 1996) that sublimate at low temperature conditions and (iii) to lack aqueous altered minerals like phyllosilicates (Brownlee et al., 2012) or to only feature them as exogenic contamination (Oklay et al., 2016). Combined this suggests that water ice never melted on these objects\footnote[1] {A possible explanation for the lack of evidence for H$_2$O at Kuiper belt object Arrokoth involves radiolysis of mixed H$_2$O and CH$_4$ ices, where CH$_4$ and H$_2$O could be consumed (Grundy et al., 2020).}. On the other hand icy planetesimals probably formed within the first few million years after the start of the solar system, thus it can be expected that short-lived radioactive isotopes like $^{26}$Al were present in their interiors and caused an internal temperature increase (e.g. Prialnik et al., 1987; Prialnik and Podolak, 1995; Merk and Prialnik, 2003, 2006). 
Additionally, collisional evolution could have led to significant heating. The most accurate models of solar system dynamical evolution (Nesvorný and Morbidelli, 2012; Nesvorný, 2015a, 2015b; Nesvorný and Vokrouhlický, 2016; Nesvorný et al., 2017) suggest that the initial trans-Neptunian icy planetesimal disk had time to evolve to collisional equilibrium before being dispersed. This means that present-day comet-size objects (typically a few km in diameter) should have experienced many collisions and most of those we see today should be fragments of originally larger objects (Weissman et al., 2020). For relatively small, kilometer-size objects, Jutzi and Benz (2017) and Schwartz et al. (2018) found that only a small fraction of the largest remnants resulting from disruptive collisions (ranging from sub-catastrophic to catastrophic) would have been heated by more than a few degrees. However, collisional disruptions of parent bodies with larger radii ($R$ = 25-50 km) lead to more significant heating (Jutzi and Michel, 2020).

Whether or not such a combined thermal and collisional evolution would have preserved the observed primitive properties of present-day comets remains to be clarified. For this purpose here we study under which conditions comets lost their highly volatile ices like CO, CO$_2$ and NH$_3$, when considering the effects of both short-lived radiogenic heating and impact heating on their precursor bodies. Here we focus on large ($R$ = 10-50 km) objects and investigate the maximum effect both of these heat sources could have on icy planetesimals to estimate under which conditions comets can remain primitive. For this purpose we perform both geodynamical thermal evolution models and SPH collision simulations.\\

Two scenarios of comet formation have been suggested in recent years: A hierarchical accretion of dust and ice grains occurring in the protoplanetary disk that took up to 25 million years (e.g. Davidsson et al., 2016) or alternatively that comets formed due to streaming instabilities followed by gravitational collapse (Wahlberg Jansson and Johansen, 2014, 2017; Wahlberg Jansson et al., 2017) that occurs on timescales ranging from $<10^2$ years (Nesvorný et al., 2021) to $10^3$ years (Wahlberg Jansson and Johansen, 2014). In this study we consider the latter formation mechanism for our thermal evolution calculations, thus estimating the upper temperature limit since an extended hierarchical accretion would allow for efficient cooling of the accreting icy planetesimals.\\
The overall goal is to provide constraints for the initial size and formation time of comets (or their precursors) formed directly by gravitational collapse.  


\section{Numerical methods}
\label{sec:methods}

\subsection{Governing equations of the thermal evolution model}
\label{sec:thermoequations}
We study the early thermal evolution of instantaneously formed icy planetesimal using the finite-difference marker-in-cell code I2ELVIS (Gerya and Yuen, 2007). The code solves the fluid dynamic conservation equations using the extended Boussinesq approximation in 2D infinite cylinder geometry. For this purpose we solve the conservation equations on a fully staggered-grid (Gerya and Yuen, 2003), namely the mass conservation equation
\begin{align}
\frac{\partial \rho}{\partial t} + \nabla \rho \mathbf{u} = 0,
\end{align}
with density $\rho$, time $t$ and flow velocity $\mathbf{u}$; the momentum conservation equation
\begin{align}
\nabla \mathbf{\sigma'} - \nabla P + \rho \mathbf{g} = 0,
\end{align}
where $\mathbf{\sigma}'$ is the deviatoric stress tensor, $P$ gives the pressure and the directional gravity $\mathbf{g}$ is obtained from the location-dependent Poisson equation
\begin{align}
\nabla^2 \Phi = 4 \pi G \rho,
\end{align}
where $\Phi$ is the gravitational potential and $G$ is Newton's constant.
Finally the energy equation is given as:
\begin{align}
\rho c_{\mathrm{P}} \left( \frac{\partial T}{\partial t} + u_{\mathrm{i}} \cdot \nabla T \right) = - \frac{\partial q_{\mathrm{i}}}{\partial x_{\mathrm{i}}} + H_{\mathrm{r}} + H_{\mathrm{s}} + H_{\mathrm{L}},
\label{eq:energy-conservation}
\end{align}
with heat capacity $c_{\mathrm{P}}$, temperature $T$, heat flux $q_{\mathrm{i}} = -k \frac{\partial T}{\partial x_{\mathrm{i}}}$, thermal conductivity $k$, spatial coordinate $x_i$, and radiogenic ($H_{\mathrm{r}}$), shear ($H_{\mathrm{s}}$) 
and latent ($H_{\mathrm{L}}$) heat terms. For iron/silicate mixtures we assume that olivine controls the rheology and the physical parameters except density. The water ice phase is considered to be pure ice I. For both materials we employ a temperature- and pressure-dependent rheology (Ranalli, 1995; Durham and Stern, 2001).

\subsubsection{Heating by short-lived radionuclides}
Due to the small size of the icy planetesimals considered in this study only the most powerful radioactive isotopes are relevant for their thermal evolution. 
In the early solar system the radioactive isotopes $^{26}$Al and $^{60}$Fe are dominant, while long-lived ones like $^{40}$K, $^{235}$U, $^{238}$U and $^{232}$Th can be neglected (Prialnik et al., 1987). 
Here we consider time-dependent radiogenic heating by both $^{26}$Al and $^{60}$Fe with half-life times of 0.716 Myr (Barr and Canup, 2008) and 2.62 Myr (Rugel et al., 2009), respectively.
For the initial $^{26}$Al/$^{27}$Al ratio we adopt a value of $5.25 \cdot 10^{-5}$ at calcium-aluminium-rich inclusion (CAI) formation (Kita et al., 2013), while for $^{60}$Fe we use an initial ratio of $^{60}$Fe/$^{56}$Fe = $1.9 \cdot 10^{-7}$ (Tachibana and Huss, 2003; Mostefaoui et al., 2004). However it should be pointed out that the $^{26}$Al/$^{27}$Al and $^{60}$Fe/$^{56}$Fe values for comets and KBOs are so far unconstrained (Prialnik et al. 2008; Merk and Prialnik, 2006; Choi et al. 2002; Guilbert-Lepoutre et al. 2019).\\

\subsubsection{Effect of porosity}
We model the icy planetesimals as a mixture of highly porous water ice and diapirs consisting of silicate and iron (see section \ref{sec:initial_conditions}).
Porosity affects the density, thermal conductivity and heat capacity of all materials considered.
The density of all components is assumed to be dependent on both temperature $T$ and porosity $\phi$.
The temperature-dependent part of the density equations for iron and silicates is described by
\begin{align}
\rho_{fe} = \rho_{fe0} \cdot [1-\alpha_{fe}(T-T_0)]
\end{align}
and
\begin{align}
\rho_{si} = \rho_{si0} \cdot [1-\alpha_{si}(T-T_0)]
\end{align}
where $\rho_{fe0}$ and $\rho_{si0}$ are the reference densities of both iron and silicates at reference temperature $T_0$, while $\alpha_{fe}$ and $\alpha_{Si}$ are the thermal expansivities for iron and silicates, respectively.
The effective density of the iron/silicate mixture constituting the diapirs is given as
\begin{align}
\rho_{Diap} = [v\rho_{fe}+(1-v)\rho_{Si}] \cdot (1-\phi)
\end{align}
where $v$ is the volume fraction of the iron and $\phi$ is the porosity of the iron/silicate mixture.
The density of the water ice material is given as
\begin{align}
\rho_{ice} = \rho_{ice0} \cdot [1-\alpha_{ice}(T-T_0)] \cdot (1-\phi)
\end{align}
where $\rho_{ice0}$ is the reference density of the water ice and $\alpha_{ice}$ is its thermal expansivity.
For the heat capacity we assume for all considered materials that it is only porosity-dependent (Davidsson et al., 2016):
\begin{align}
c_P = c_{P0} \cdot (1-\phi)
\end{align}
where $c_{P0}$ is the reference value of the heat capacity for the specific material.

The effective thermal conductivity $k_{eff}$ for highly porous material ($\phi > 0.4$) as used in our study is described as
\begin{align}
k_{eff} = k \cdot e^{a-\phi/\phi_{\mathrm{1}}}
\end{align}
with constants $a=-1.2$ and $\phi_{\mathrm{1}} = 0.167$, using an approximation based on laboratory experiments (Henke et al., 2012; Gail et al., 2015).\\ 
The physical parameters for both non-porous solid silicates and solid iron are taken from the compilation in Lichtenberg et al. (2018). The physical parameters for non-porous ice I and water are taken from Kirk and Stevenson (1987).\\
Due to the low internal pressures and temperatures well below the sintering temperature (Yomogida and Matsui, 1984), no change of porosity is expected for the iron/silicate diapirs. On the other hand water ice melting would reduce the porosity of this material to zero. For simplicity we do not treat the change of ice porosity in these calculations.

\subsubsection{Treatment of highly volatile ices}
\label{sec:loss-model}
Highly volatile ices like CO, CO$_2$ and NH$_3$ are not directly considered in our model. We track the losses of highly volatile ices occurring at 40 K (CO), 80 K (CO$_2$) and 140 K (NH$_3$) (Davidsson et al., 2016) by determining the number of ice markers that have overstepped the temperature criteria over time. Assuming grain sizes as observed at the Philae landing site on 67P/Churyumov-Gerasimenko (Poulet et al., 2016) and considering the porosity values in our models (see section \ref{sec:initial_conditions}), Darcy's law (Darcy, 1856) can be used to estimate a characteristic segregation time scale (Lichtenberg et al., 2019a) for the gas loss of $<$1 year. Based on this estimation we assume that the small abundance (relative to H$_2$O) of a specific highly volatile ice is instantaneously lost to space from a certain marker after its sublimation temperature has been overstepped. This provides a lower limit on the preservation of highly volatile ices.\\

\subsubsection{Water ice melting model}
\label{sec:melt-model}
Due to the small size of objects like icy planetesimals the pressure-effect on the melting temperature of pure ice I is minor, thus we use as water ice melting criteria a temperature of 273 K and a pressure above 600 Pa (Prialnik et al., 2008) and assume a latent heat of 284 kJ/kg (Kirk and Stevenson, 1987; Grasset et al., 2000). While in reality the viscosity of water is $\approx10^{-3}$ Pa s (Kestin et al., 1978), due to numerical limitations we assume a water viscosity of $10^{12}$ Pa s, corresponding to the lower cut-off viscosity (see also section 2.2). In order to approximate the increased heat loss caused by the presence of convecting low viscosity water, we increase the effective thermal conductivity of the water as described and used in previous studies (Tackley et al., 2001; Hevey and Sanders, 2006; Golabek et al., 2011; Lichtenberg et al., 2018). Both the latent heat and the more efficient heat loss tend to stabilize the water temperature close to the melting temperature. It should be noted that we ignore the potential influence of highly volatile ices on the melting temperature of ice I. Since we are mostly interested in the evolution of icy planetesimals at low temperatures, we do not consider here the aqueous alteration of silicates (Lichtenberg et al., 2019b) and potential mud convection (Bland and Travis, 2017).\\

\subsection{Initial conditions}
\label{sec:initial_conditions}
In all I2ELVIS calculations the model box has dimensions of 120$^2$ km, resolved by 501 grid points in both directions, thus corresponding to a grid resolution of 240 m. Each grid cell is initially filled with 16 markers storing all physical properties.
The icy planetesimals are assumed to be circular with radii ranging from 10 to 50 km, the rest of the numerical domain is filled with low viscosity sticky air material establishing a free surface (Schmeling et al., 2008; Crameri et al., 2012). In all calculations we fix the icy planetesimal density at $\approx470$ kg/m$^3$, corresponding to typical densities observed for present-day comets (Sierks et al., 2015) and Kuiper belt objects like (486958) Arrokoth (McKinnon et al., 2020; Spencer et al., 2020), while considering rock-to-ice mass fractions $f$ ranging from $1/3$ to $4/5$, spanning roughly the range of values used in planet formation models or deduced from observations (e.g. Marboeuf et al., 2014; Davidsson et al., 2016).
To numerically resolve the regions contributing to radiogenic heating inside the icy planetesimal for simplicity we assume that ice is separated from the silicates and the iron, while in reality all phases could be mixed on the small scale (see also 5. Model limitations). For this purpose we define circular diapirs that contain a volume fraction of iron of $v = 0.1$ and silicates make up 90 \% of the diapir, while the remainder of the icy planetesimal interior is made of water ice.\\
Each of these iron/silicate diapirs has a radius of 720 m and is heated by both $^{26}$Al and $^{60}$Fe, while no radioactive isotopes are present in the ice I. We tested different radii of the iron/silicate diapirs and found comparable results, thus use the radius of 720 m for the remainder of this study. To simulate different rock-to-ice mass fractions $f$, we vary the number of iron/silicate diapirs that are randomly positioned inside the respective icy planetesimal. For example in a $R$ = 50 km icy planetesimal the number of iron/silicate diapirs can range from 522 to 2376, depending on the rock-to-ice mass fraction $f$ considered. These diapirs are allowed to touch each other, but not to overlap. To avoid an increase of the mean density of the constructed icy planetesimal, we increase the porosity of both iron/silicate diapirs and ice with increasing rock-to-ice mass fraction $f$. The used porosities $\phi$ range from 0.63 for models with $f$ = $1/3$ to 0.805 for a rock-to-ice mass fraction of $4/5$. Since the potential energy release related to the instantaneous formation of small objects 
like icy planetesimals is negligible (Schubert et al., 1986), we use for all of our simulations the assumption that both space and initial temperature $T_{start}$ of the icy planetesimal are identical. 
The temperature of space, here given by the surrounding sticky air material is fixed at $T_{start}$, thus representing an infinite reservoir absorbing heat released from the icy planetesimal (Golabek et al., 2011). 
For all calculations we use for the viscosity cut-offs at 10$^{12}$ and 10$^{24}$ Pa s.\\

Based on the outcome of the thermal evolution models, we determine the maximum temperature reached at various depths within the icy planetesimals while assuming radial symmetry. These results are then used together with the results of the SPH collision simulations (see section \ref{sec:collisionmodel}) to estimate whether interior thermal evolution or collisional heating is dominant for certain regions inside specific icy planetesimals.

\subsection{Parameter space}
\label{sec:parameterspace}
We perform 268 thermal evolution simulations in order to cover the parameter space. Merouane et al. (2016) suggests that icy planetesimals with $R \geq$ 50 km would compact due to sufficiently high internal pressures, thus we use this as upper limit for the radius, this way still covering objects like C/1995 O1 Hale-Bopp ($R \approx$ 35 km) (Morbidelli \& Nesvorný, 2019). Based on these constraints we vary the icy planetesimal radii $R$ from 10 to 50 km.
We test formation times $t_{form}$ ranging from 2.5 to 5 Myr after CAI formation. Previous studies suggest that earlier formation times $t_{form}<$ 2.5 Myr lead to very strong interior heating and pore closure (Lichtenberg et al. 2016), while for later formation times $t_{form} >$ 5 Myr, the interior heating becomes negligible. 

As additional parameter we vary the rock-to-ice mass fraction $f$ from $1/3$ to $4/5$. For most models we use an initial and space temperature of 30 K, however we test also initial temperatures of 50 and 100 K, thus covering the typical formation temperatures suggested in previous literature (Prialnik et al., 2008; Bar-Nun et al., 2013; Rubin et al., 2015a). These additional calculations with higher start temperatures display very similar results to those with a start temperature of 30 K, the only difference being an offset of the peak temperatures to higher values. Based on this, we will discuss for the remainder of the study the results with $T_{start}$ = 30 K.\\

\subsection{Collision model}
\label{sec:collisionmodel}
\subsubsection{Simulation data set}
The collision simulations and data set used in this paper are described in detail in Jutzi et al. (2019) and Jutzi and Michel (2020). To model asteroid disruptions and subsequent reaccumulations, Jutzi et al. (2019) used a SPH/N-body approach (e.g. Michel et al., 2001, 2003). The early phases of the collisions were simulated applying a SPH impact code (e.g. Jutzi, 2015) with self-gravity, various material models including strength, friction and porosity and the Tillotson equation of state (Tillotson, 1962). The N-body code pkdgrav (Richardson et al., 2000) was then used to compute the dynamical evolution of the system to late time.

Jutzi et al. (2019) used porous parent bodies with a nominal radius of $R$ = 50 km and a matrix of impact conditions covering a wide range of impact speeds (from 3 to 7 km/s), impact angles (from 15$^\circ$ to 75$^\circ$ with 15$^\circ$ increments) and impactor radii ranging from $\approx$ 3 to $\approx$ 18 km.
Here, we perform additional SPH simulations (of the impact phase only) using smaller $R$ = 20 km targets (Table \ref{table:4runs}).

\subsubsection{Heating analysis}
The computation of the impact-induced temperature increase is based on the outcome of the SPH impact simulations, as described in detail in Jutzi and Michel (2020).
To compute the temperature increase from the specific internal energy, a temperature-dependent heat capacity is applied. For this study, we use parameters corresponding to a comet-like material with a rock-to-ice mass fraction $f$ of 2/3 (Jutzi and Michel, 2020).

\subsection{Combining interior and collisional heating}
\label{sec:combined_model}
The thermal evolution simulations described in section \ref{sec:parameterspace} provide the maximum temperature $T_{max,int}$ reached at various depths within the icy planetesimals as a function of their radius and formation time.

For the combined analysis of the radiogenic and impact heating, we assume that the collision took place a long time after the formation of the body. Thus the icy planetesimal has cooled to the initial equilibrium temperature $T_{start}$ when the collision occurs. For the objects used in the SPH collision simulations we then determine for each particle, located at a given initial distance from the center $r$, the maximum temperature experienced due to the heating $T_{max,col}$ caused by the specific impact.

The overall temperature maximum is then calculated for each particle by $T_{max}$ = max($T_{max,int}$,$T_{max,col}$). Using this approach, we compute the mass fraction of material that experienced a maximum temperature $T_{max}$ larger than a critical temperature $T_{crit}$, with $T_{crit}$ = 40 K, 80 K and 140 K, respectively (as discussed in section \ref{sec:loss-model}).


\section{Results}
\label{sec:results}

\subsection{Thermal evolution}
\label{sec:therm_evo}
In general the thermal evolution of the icy planetesimals considered here lasts only for a few million years due to their small size, thus allowing for their fast cooling as found by previous studies for small objects (e.g. Prialnik et al., 1987; Lichtenberg et al., 2018). For all cases an early heating phase occurs due to the presence of especially $^{26}$Al and the central region of the icy planetesimal displays the highest temperatures. Dependent on the formation time, controlling the initial amount of $^{26}$Al, and the radius, influencing how efficient the body can cool, the interior can become warm enough to lose its complement of highly volatile ices or not. After several half-life times of $^{26}$Al this radionuclide goes extinct and the cooling of the interior starts from the surface downwards. Even for the largest bodies considered here the cooling of the deep interior starts within less than 10 Myr after CAI formation. 
As expected, the peak temperatures and the related preservation of highly volatile ices inside the icy planetesimals strongly depend on formation time $t_{\mathrm{form}}$, radius $R$ and rock-to-ice mass ratio $f$ as compiled in Fig. 1.
The results show that preservation of CO mostly depends on the formation time and radius of the icy planetesimal, not so much on the rock-to-ice mass ratio $f$ (see Fig. 1a). Significant CO preservation occurs preferentially in objects with radii $\le$ 20 km that formed later than 3.5 Myr after CAIs. On the other hand CO$_2$ preservation is viable even in early formed objects as long as these objects are sufficiently small ($R \le$ 20 km) and their rock-to-ice mass fraction is smaller than $1/2$, while for later formed objects the CO$_2$ preservation is less dependent on the radius and the rock-to-ice mass fraction (see Fig. 1b). Due to the relatively high sublimation temperature of NH$_3$ most of the considered icy planetesimals can preserve NH$_3$ in the layers near the surface. This is also true for early formed and large bodies as long as their rock-to-ice mass fraction is less than $1/2$ (see Fig. 1c). For a larger rock-to-ice mass fraction these large bodies can only preserve a near surface layer of NH$_3$. Finally water ice melting is possible in early formed ($t_{\mathrm{form}}\le$ 2.75 Myr after CAIs) and large icy planetesimals ($R \ge$ 30 km) for rock-to-ice mass fraction $\ge2/3$ (see Fig. 1d).

\subsection{Combined interior and collisional heating}
The procedure described in section \ref{sec:combined_model} is used to calculate the combined effects of interior and collisional heating. The peak temperatures reached by radiogenic heating, which are used for this analysis, are shown in Figure \ref{fig:peaktemp} as a function of radius for different icy planetesimal radii and formation times. 

To illustrate the relative effects of interior and impact heating, we display in Figures \ref{fig:heating_vs_q_t0_3ma} and \ref{fig:heating_vs_q_t0_4ma} the results of our combined analysis using $R$ = 50 km objects and two different formation times ($t_{form}$ = 3 and 4 Myr). We consider a fixed set of impact velocity and impact angle ($v_{imp}$ = 3 km/s, $\theta$ = 45$^\circ$) and compute the mass fraction heated above a certain temperature ($T_{crit}$ = 80 K) as a function of specific impact energy, normalised by the catastrophic disruption threshold $Q^*_D$ (Jutzi et al., 2019; Jutzi and Michel, 2020). As in Jutzi and Michel (2020), we distinguish between material which is bound and gets re-accreted onto the largest remnant and the unbound material ejected from the system (see Jutzi and Michel, 2020 for details). As expected, the contribution from interior heating is more significant for early formation times, and it affects more the bound than the unbound material, as the latter originates from shallower depths inside the target, which reach lower peak temperatures (see Figure \ref{fig:peaktemp}). 

For the further analysis, we consider for two different target sizes ($R$ = 20 km and $R$ = 50 km) two representative cases: a nearly catastrophic collision ($Q/Q^*_D \approx 0.44$) and a cratering impact ($Q/Q^*_D \approx 0.0044$). The corresponding parameters are given in Table \ref{table:4runs}. For each case, we compute the mass fraction that experienced a critical maximum temperature caused by either interior or impact heating. The critical temperatures are $T_{crit}$, with $T_{crit}$ = 40 K, 80 K and 140 K, respectively (see section \ref{sec:loss-model}). As above, we distinguish between material which is bound and gets re-accreted onto the largest remnant and the unbound material ejected from the system.

The result of our analysis (Figures \ref{fig:t_r_50_bound} - \ref{fig:t_r_20_unbound}) allows us to determine a critical radius $R_{crit}$ as a function of formation time for a given critical temperature $T_{crit}$. We define $R_{crit}$ such that at least 50\% of the mass (of either bound or unbound material) has never experienced $T>T_{crit}$. To determine $R_{crit}$ we use a simple linear interpolation between the $R$ = 20 km and $R$ = 50 km results shown in Figures \ref{fig:t_r_50_bound} - \ref{fig:t_r_20_unbound} and compute the radius at which the combined heating curve equals a mass fraction of $f_{heated}$ = 0.5 (Figure \ref{fig:rcrit_tform}). To check the influence of our choice on the results, we also considered smaller and larger values of $f_{heated}$ (Figure \ref{fig:rcrit_tform_fractions}). For comparison, we also show in these figures the corresponding curves for interior heating only, thus without considering impact heating.  

These calculations allow for an estimate of the maximum initial parent body size of a specific comet for a given formation time, assuming that it formed in a given collision scenario (catastrophic disruption or cratering event) and that it has never reached $T>T_{crit}$ (see discussion in section \ref{sec:discussion}). 


\section{Discussion}
\label{sec:discussion}
The results of our study allow for constraints regarding the initial radius and formation time of icy planetesimals (the precursors of comets and small KBOs), as illustrated in Figure \ref{fig:rcrit_tform}. Below we discuss some examples, however we note that the following three cases are just illustrative examples of cratering and catastrophic collision events that may have lead to objects like 67P/Churyumov-Gerasimenko, C/1995 O1 Hale-Bopp and (486958) Arrokoth. One has to keep in mind that the actual heating effects in the given regimes somewhat vary as a function of specific impact energy (Figures \ref{fig:heating_vs_q_t0_3ma} and \ref{fig:heating_vs_q_t0_4ma}).
The results also depend on the assumed fraction of material heated above a critical temperature  $f_{heated}$ (Figure \ref{fig:rcrit_tform_fractions}) and here we assume $f_{heated}$ = 0.5.

\subsection{Application to 67 P/Churyumov-Gerasimenko}
Comet 67 P/Churyumov-Gerasimenko contains substantial amounts of volatiles (Rubin et al., 2020), which set constraints for its formation and history (e.g. Davidsson et al., 2016). 
Based on our results, we discuss various scenarios for its formation by a collisional event involving a larger parent body. The case in which comet 67 P/Churyumov-Gerasimenko is assumed to be a primordial object is not discussed here, because its size is not within the range investigated in this study.

Assuming a critical temperature of $T_{crit}=$ 40 K, the parent body of comet 67 P/Churyumov-Gerasimenko could not have been larger than $R_{crit} \approx$ 20 km, if the comet resulted from the unbound ejecta of a cratering impact (blue line in Figure \ref{fig:rcrit_tform}d).
Assuming the parent body formed at $t_{form}$ $\approx$ 5 Myr and experienced a catastrophic collision, the comet could have originated from the largest remnant (which was subsequently disrupted). In this case the original parent body must have been smaller than $R \approx$ 40 km (red line in Figure \ref{fig:rcrit_tform}a). If the comet originated from the unbound material, then the original parent body must have been even smaller, $R_{crit} \lessapprox$ 15 km (blue line in Figure \ref{fig:rcrit_tform}a). The earlier the formation time, the smaller are the critical radii and no solution exists in these scenarios for formation times $<$ 4.25 Myrs (origin from largest bound remnant) and 4.65 Myrs (origin from unbound material) for the $R >$ 10 km parent bodies considered here.         

On the other hand, for a much higher critical temperature of $T_{crit}=$ 140 K (which is probably unrealistic, given the abundance of highly volatile ices), the parent body of comet 67 P/Churyumov-Gerasimenko could have been as large as $R_{crit} \approx$ 70 km, if $t_{form}$ $>$ 3.5 Myr and the comet resulted from the unbound material of a catastrophic disruption (blue line in Figure \ref{fig:rcrit_tform}c). The parent body could have been even larger if the comet originated from the bound material (red line in Figure \ref{fig:rcrit_tform}c).

\subsection{Application to C/1995 O1 Hale-Bopp}
Large objects such as comet C/1995 O1 Hale-Bopp with a radius of $\approx$ 35 km (Morbidelli \& Nesvorný, 2019) could not have been formed as ejecta from a cratering event. If C/1995 O1 Hale-Bopp originated from the bound material of a catastrophic disruption, its parent body radius would have been $\approx$ 44 km, assuming the same density for both objects.
Observations of C/1995 O1 Hale-Bopp show no evidence of significant depletion of CO (A'Hearn et al., 2012; Womack et al., 2017, 2021). We therefore consider a critical temperature of $T_{crit}$ = 40 K, for which this scenario requires formation times $>$ 5 Myrs according to our results (Figure \ref{fig:rcrit_tform}a). 

If Hale-Bopp formed from the unbound ejecta of a catastrophic disruption of an object of $\approx$ 10 times its mass, its parent body would have had to have a radius of $\approx$ 75 km. This scenario is not feasible according to our model.

If comet C/1995 O1 Hale-Bopp is assumed to be a primordial object, the required formation times are somewhat earlier than in the scenario involving the disruption of larger object as described above, because of the smaller initial size (35 km vs. 44 km) and because there is slightly less heating than in the case of a collisional disruption (black dotted vs. red line in Figures \ref{fig:rcrit_tform}a-c). For a critical temperature of $T_{crit}$ = 40 K, as suggested by the observation of CO outgassing, the primordial scenario requires formation times  $>$ 4.75 Myrs.

\subsection{Application to (486958) Arrokoth}
The sublimation evolution of New Horizons target (486958) Arrokoth has recently been investigated (Lisse et al., 2021; Steckloff et al., 2021; Zhao et al., 2021). Here we discuss possible constraints from our analysis of combined impact and interior heating.  

The two lobes of (486958) Arrokoth have equivalent radii of $\approx 8$ and $\approx 6.5$ km, combined in one object this yields an equivalent radius of $\approx 9.5$ km (Spencer et al., 2020). If (486958) Arrokoth originated from the bound material of a catastrophic disruption, its parent body radius would have been $\approx$ 12 km, assuming the same density for both objects. Although these radii are at the lower limit of the size range considered in our study, our analysis shows that the bulk of (486958) Arrokoth never experienced significant heating ($T > $ 80 K). This confirms that Arrokoth’s interior temperature was never high enough for amorphous ice to crystallize and expel its payload of highly volatile ices (Grundy et al., 2020). This is the case when assuming that (486958) Arrokoth formed in a catastrophic disruption, but also in case (486958) Arrokoth is assumed to be a primordial object, as the maximum temperatures in the latter case would be even slightly lower (Figure \ref{fig:rcrit_tform}). 

\subsection{Dependence on fraction of heated material $f_{heated}$}
The maximum radii and formation times discussed in the three cases above depend on the assumed fraction of heated material $f_{heated}$. Figure \ref{fig:rcrit_tform_fractions} compares the results obtained with our nominal $f_{heated} = 0.5$ with the cases using $f_{heated}$ = 0.1 and 0.9, respectively. With increasing (decreasing) $f_{heated}$, the maximum radii become larger (smaller) and the possible formation times become smaller (larger).


\section{Model limitations}
\label{sec:limitations}
In the I2ELVIS simulations we ignore the potential sintering of ice grains at higher temperatures that might affect density, thermal conductivity and heat capacity since we are mostly interested in losses of highly volatile ices occurring at rather low temperatures. Kuiper belt object (66652) Borasisi is larger than the icy planetesimals considered in our study, however its high density (Vilenius et al., 2014) suggests that loss of porosity due to ice melting can occur, thus this effect should be considered in future work. Also the highly volatile ices are only considered via temperature thresholds, ignoring their actual sublimation and outgassing. Related we ignore the latent heat associated to the sublimation of these highly volatile ices. In both the I2ELVIS and the SPH models so far we assume that the small abundances of highly volatile ices are instantaneously lost to space when reaching their sublimation temperature. In reality these gases can diffuse both upward toward the surface and inward to greater depths and can potentially deposit in cold regions (Espinasse et al., 1991) - for the I2ELVIS models this could be the cold surface, while for the SPH models this could be regions that remain cold. Additionally it should be kept in mind that the gases will in reality carry some heat away (Prialnik et al., 2004) that we do not consider so far.\\

The analysis of the collisional outcome is based on the SPH impact simulation, and we only distinguish between the largest remnant and material not bound to this remnant. The properties of the individual smaller fragments will be investigated in a future study using coupled SPH/N-body simulations (as done by Schwartz et al. (2018) and Michel et al. (2020)). Computing the size distribution will then also allow for an investigation of the temperature evolution of the individual fragments, which is neglected in this study. However, we point out that results show that for the largest remnant (i.e. the bound material), the interior heating is more significant than the impact heating. Therefore, the details of the post-impact temperature evolution and the heating/cooling of surrounding material do not affect the overall results (combined heating effects) in this case. 

In the calculations of the combined effects of interior and collisional heating, we use impact simulation results based on a fixed impact velocity (3 km/s). However, we note that the degree of impact heating of the escaping material does not strongly depend on impact velocity for a given specific impact energy normalized by $Q^*_D$ (see Jutzi and Michel, 2020, Figure 2). 

The critical maximum radius (Figure \ref{fig:rcrit_tform}) is computed based on simulations using two target radii and two impact regimes only. The radii resulting from the applied linear interpolations are therefore only approximately valid in a limited size range (10 km $\lessapprox$ $R$ $\lessapprox$ 100 km).
Finally, we assume that the collision took place a long time after the formation of the body and that the icy planetesimal has cooled to the initial equilibrium temperature $T_{start}$. If the collision happened very early on, the interior temperatures $T>T_{start}$ (which depend on the formation time as well as the time of the collision) could have lead to an increase of the maximal temperatures reached by the collisional heating.


\section{Conclusions \& Outlook}
\label{sec:conclusions}
Although the results of our analysis - such as the potential maximum initial radii of icy planetesimals depend on a number of assumptions such as the critical temperatures $T_{crit}$ and fractions of heated materials $f_{heated}$, there are also some robust trends. We find that for the part of the material which is bound to the largest remnant after the collision, the interior heating is more significant than the impact heating. For the formation times considered here (2.5-5 Myrs), the maximum temperatures that this material has  experienced is mainly given by the interior heating. Therefore, it is indeed the formation time which determines the degree of material processing in this case. This is true for the whole range of impact energies considered, from cratering impacts to catastrophic disruptions. 
From this point of view, it is therefore difficult to distinguish between the scenario of primordial formation of comets (and/or small KBOs) and the scenario of their formation by a collisional event such as a catastrophic disruption (where the resulting object corresponds to the largest remnant). 

However, for the unbound material produced in collisions, the impact heating is much more significant (as already found by Jutzi and Michel, 2020) and, except for very early formation times, dominates over interior heating. Therefore, for objects resulting from such unbound material, the initial size of the parent body is more important in terms of heating than its formation time. In such scenarios, the maximum initial size of the parent body can in principle be constrained independently of the formation time. 

However, in all scenarios, crucial quantities are the critical temperature $T_{crit}$ and the fraction of the material $f_{heated}$ of a given object that has experienced temperatures $T > T_{crit}$. Future observations and missions may be able to better constrain these quantities.


\section*{Acknowledgements}
\label{sec:acknowledgements}
We thank two anonymous reviewers for constructive comments that helped to improve the manuscript considerably. The authors thank Taras V. Gerya for providing the code I2ELVIS. We thank Tim Lichtenberg for helpful comments. Computations have been performed on clusters btrzx2, University of Bayreuth and Horus, University of Bern. The perceptually-uniform color map \textit{roma} used in Fig. 1 is taken from Crameri (2018). M.J. acknowledges support from the Swiss National Centre of Competence in Research PlanetS.

\pagebreak


\begingroup
\section*{References}
\noindent
A'Hearn, M.F. and 20 colleagues, 2012. Cometary volatiles and the origin of comets. Astrophys. J. 758, 29.\\
Bar-Nun, A., Laufer, D., Rebolledo, O., Malyk, S., Reisler, H., Wittig, C., 2013. Gas Trapping in Ice and Its Release upon Warming. In: The Science of Solar System Ices, M. S. Gudipati, J. Castillo-Rogez (Eds.), 487–499, Springer, New York.\\
Barr, A. C., Canup, R. M., 2008. Constraints on gas giant satellite formation from the interior states of partially differentiated satellites. Icarus 198, 163–177.\\
Biver, N. et al., 1996. Substantial outgassing of CO from comet Hale-Bopp at large heliocentric distance. Nature 380, 137-139.\\
Bland, P. A., Travis, B. J., 2017. Giant convecting mud balls of the early solar system. Sci. Adv. 3, e1602514.\\
Brownlee, D., Joswiak, D., Matrajto, G., 2012. Overview of the rocky component of Wild 2 comet samples: Insight into the early solar system, relationship with meteoritic materials and the differences between comets and asteroids. Meteorit. Planet. Sci. 47, 453–470.\\
Choi, Y.-J., Cohen, M., Merk, R., Prialnik, D., 2002. Long-term evolution of objects in the Kuiper belt zone - effects of insolation and radiogenic heating. Icarus 160, 300–312.\\
Crameri, F., Schmeling, H., Golabek, G. J., Duretz, T., Orendt, R., Buiter, S. J. H., May, D. A., Kaus, B. J. P., Gerya, T. V., Tackley, P. J., 2012. A comparison of numerical surface topography calculations in geodynamic modelling: an evaluation of the ‘sticky air’ method. Geophys. J. Int. 189, 38–54.\\
Crameri, F., 2018. Geodynamic diagnostics, scientific visualisation and StagLab 3.0. Geosci. Model Dev. 11, 2541–2562.\\
Darcy, H. P. G., 1856. Les Fontaines publiques de la ville de Dijon: exposition et application des principes à suivre et des formules à employer dans les questions de distribution d’eau. V. Dalamont, Paris.\\
Davidsson, B. J. R. et al., 2016. The primordial nucleus of comet 67P/Churyumov-Gerasimenko. Astron. Astrophys. 592, A63.\\
Durham, W. B., Stern, L. A., 2001. Rheological properties of water ice - Applications to satellites of the outer planets. Annu. Rev. Earth Planet. Sci. 29, 295–330.\\
Edgeworth, K. E., 1949. The origin and the evolution of the Solar System. Mon. Not. R. Astron. Soc. 109, 600-609.\\
Espinasse, S., Klinger, J., Ritz, C., Schmitt, B., 1991. Modeling of the thermal behavior and of the chemical differentiation of cometary nuclei. Icarus 92, 350–365.\\
Gail, H.-P., Henke, S., Trieloff, M., 2015. Thermal evolution and sintering of chondritic planetesimals. II. Improved treatment of the compaction process. Astron. Astrophys. 576, A60.\\
Gerya, T. V., Yuen, D. A., 2003. Characteristics-based marker-in-cell method with conservative finite-differences schemes for modeling geological flows with strongly variable transport properties. Phys. Earth Planet. Inter. 140, 293–318.\\ 
Gerya, T. V., Yuen, D. A., 2007. Robust characteristics method for modelling multiphase visco-elasto-plastic thermo-mechanical problems. Phys. Earth Planet. Inter. 163, 83–105.\\
Golabek, G. J., Keller, T., Gerya, T. V., Zhu, G., Tackley, P. J., Connolly, J. A. D., 2011. Origin of the martian dichotomy and Tharsis from a giant impact causing massive magmatism. Icarus 215, 346–357.\\
Grasset, O., Sotin, C., Deschamps, F., 2000. On the internal structure and dynamics of Titan. Planet. Space Sci. 48, 617-636.\\
Grundy, W. M. et al., 2020. Color, composition, and thermal environment of Kuiper Belt object (486958) Arrokoth. Science 367, eaay3705.\\
Guilbert-Lepoutre, A., Prialnik, D., Métayera, R., 2019. Internal structure and cryovolcanism on Trans-Neptunian objects. In: The Trans-Neptunian Solar System, D. Prialnik, M. A. Barrucci, L. A. Young (Eds.), 183-201, Elsevier, Amsterdam.\\
Henke, S., Gail, H.-P., Trieloff, M., Schwarz, W.H., Kleine, T., 2012. Thermal history modelling of the H chondrite parent body. Astron. Astrophys. 545, A135.\\
Hevey, P. J., Sanders, I. S., 2006. A model for planetesimal meltdown by $^{26}$Al and its implications for meteorite parent bodies. Meteorit. Planet. Sci. 41, 95–106.\\
Jutzi, M., 2015. SPH calculations of asteroid disruptions: The role of pressure dependent failure models. Planet. Space Sci. 107, 3-9.\\
Jutzi, M., Benz, W., 2017. Formation of bi-lobed shapes by sub-catastrophic collisions. A late origin of comet 67P's structure. Astron. Astrophys. 597, A62.\\
Jutzi, M., Michel, P., Richardson, D. C., 2019. Fragment properties from large-scale asteroid collisions: I: Results from SPH/N-body simulations using porous parent bodies and improved material models. Icarus 317, 215-228.\\
Jutzi, M., Michel, P., 2020. Collisional heating and compaction of small bodies: Constraints for their origin and evolution. Icarus 350, 113867.\\
Kestin, J., Sokolov, M., Wakeham, W. A., 1978. Viscosity of liquid water in the range -8 $^\circ$C to 150 $^\circ$C. J. Phys. Chem. Ref. Data 7, 941-948.\\
Kirk, R. L., Stevenson, D. J., 1987. Thermal Evolution of a Differentiated Ganymede and Implications for Surface Features. Icarus 69, 91-134.\\
Kita, N. T., Yin, Q.-Z., MacPherson, G. J., Ushikubo, T., Jacobsen, B., Nagashima, K., Kurahashi, E., Krot, A. N., Jacobsen, S.B., 2013. $^{26}$Al-$^{26}$Mg isotope systematics of the first solids in the early solar system. Meteorit. Planet. Sci. 48, 1383–1400.\\
Kuiper, G. P., 1951. On the origin of the Solar System. In: Astrophysics: A Topical Symposium, J. A. Hynek (Ed.), 357-424, McGraw-Hill, New York.\\
Lichtenberg, T., Golabek, G. J., Gerya, T. V., Meyer, M. R., 2016. The effects of short-lived radionuclides and porosity on the early thermo-mechanical evolution of planetesimals. Icarus 274, 350-365.\\
Lichtenberg, T., Golabek, G. J., Dullemond, C. P., Sch\"onb\"achler, M., Gerya, T. V., Meyer, M. R., 2018. Impact splash chondrule formation during planetesimal recycling. Icarus 302, 27-43.\\
Lichtenberg, T., Keller, T., Katz, R. F., Golabek, G. J., Gerya, T. V., 2019a. Magma ascent in planetesimals: Control by grain size. Earth Planet. Sci. Lett. 507, 154-165.\\
Lichtenberg, T., Golabek, G. J., Burn, R., Meyer, M. R., Alibert, Y., Gerya, T. V., Mordasini, C. A., 2019b. A water budget dichotomy of rocky protoplanets from $^{26}$Al-heating. Nat. Astron. 3, 307-313.\\
Lisse, C.~M. and 32 colleagues, 2021.\ On the Origin and Thermal Stability of Arrokoth's and Pluto's Ices. Icarus 356, 114072.\\
Marboeuf, U., Thiabaud, A., Alibert, Y., Cabral, N., Benz, W., 2014. From planetesimals to planets: volatile molecules. Astron. Astrophys. 570, A36.\\
McKinnon, W. B. et al., 2020. The solar nebula origin of (486958) Arrokoth, a primordial contact binary in the Kuiper Belt. Science 367, eaay6620.\\
Merk, R., Prialnik, D., 2003. Early thermal and structural evolution of small bodies in the trans-Neptunian zone. Earth Moon Planets 92, 359–374.\\
Merk, R., Prialnik, D., 2006. Combined modeling of thermal evolution and accretion of trans-neptunian objects—Occurrence of high temperatures and liquid water. Icarus 183, 283–295.\\
Merouane, S. et al., 2016. Dust particle flux and size distribution in the coma of 67P/Churyumov-Gerasimenko measured in situ by the COSIMA instrument on board Rosetta. Astron. Astrophys. 596, A87.\\
Michel, P., Benz, W., Tanga, P., Richardson, D. C., 2001. Collisions and Gravitational Reaccumulation: Forming Asteroid Families and Satellites. Science 294, 1696.\\
Michel, P., Benz, W., Richardson, D. C., 2003. Disruption of fragmented parent bodies as the origin of asteroid families. Nature 421, 608-611.\\
Michel, P., Ballouz, R.L., Barnouin, O.S., Jutzi, M., Walsh, K.J., May, B.H., Manzoni, C., Richardson, D.C., Schwartz, S.R., Sugita, S., Watanabe,S., Miyamoto, H., Hirabayashi, M., Bottke, W.F., Connolly, H.C., Yoshikawa, M., Lauretta, D.S., 2020. Collisional formation of top-shaped asteroids and implications for the origins of Ryugu and Bennu. Nat. Commun. 11, 2655.\\
Morbidelli, A., Nesvorný, A., 2019. Kuiper belt: Formation and evolution, In: The Trans-Neptunian Solar System, D. Prialnik, M. A. Barrucci, L. A. Young (Eds.), 25-59, Elsevier, Amsterdam.\\
Mostefaoui, S., Lugmair, G. W., Hoppe, P., 2005. $^{60}$Fe: A heat source for planetary differentiation from a nearby supernova explosion. Astrophys. J. 625, 271-277.\\
Nesvorný, D., 2015a. Jumping Neptune can explain the Kuiper belt kernel. Astron. J. 150, 68.\\
Nesvorný, D., 2015b. Evidence for slow migration of Neptune from the inclination distribution of Kuiper belt objects. Astron. J. 150, 73.\\
Nesvorný, D., Morbidelli, A., 2012. Statistical study of the early solar system’s instability with four, five, and six giant planets. Astron. J. 144, 117.\\
Nesvorný, D., Vokrouhlický, D., 2016. Neptune’s orbital migration was grainy, not smooth. Astrophys. J. 825, 94.\\
Nesvorný, D., Vokrouhlický, D., Dones, L., Levison, H.F., Kaib, N., Morbidelli, A., 2017. Origin and evolution of short-period comets. Astrophys. J. 845, 27.\\
Nesvorný, D., Li, R., Simon, J. B., Youdin, A. N., Richardson, D. C., Marschall, R., Grundy, W. M., 2021. Binary Planetesimal Formation from Gravitationally Collapsing Pebble Clouds. Planet. Sci. J. 2, 27.\\
Oklay, N. et al., 2016. Variegation of comet 67P/Churyumov-Gerasimenko in regions showing activity. Astron. Astrophys. 586, A80.\\
Oort, J. H., 1950. The Structure of the Cloud of Comets Surrounding the Solar System and a Hypothesis Concerning its Origin. Bull. Astron. Inst. Neth. 11, 91–110.\\
Poulet, F., Lucchetti, A., Bibring, J.-P., Carter, J., Gondet, B., Jorda, L., Langevin, Y., Pilorget, C., Capanna, C., Cremonese, G., 2016. Origin of the local structures at the Philae landing site and possible implications on the formation and evolution of 67P/Churyumov–Gerasimenko. Mon. Not. R. Astron. Soc. 462, S23–S32.\\
Prialnik, D., Bar-Nun, A., Podolak, M., 1987. Radiogenic heating of comets by $^{26}$Al and implications for their time of formation. Astrophys. J. 319, 993-1002.\\
Prialnik, D., Podolak, M., 1995. Radioactive heating of porous comet nuclei. Icarus 117, 420-430.\\
Prialnik, D., Benkhoff, J., Podolak, M., 2004. Modeling the Structure and Activity of Comet Nuclei. In: Comets II, M. C. Festou, H. U. Keller, and H. A. Weaver (Eds.), 359–387, Univ. Arizona Press, Tucson.\\
Prialnik, D., Sarid, G., Rosenberg, E. D., Merk, R., 2008. Thermal and Chemical Evolution of Comet Nuclei and Kuiper Belt Objects. Space Sci. Rev. 138, 147–164.\\
Ranalli, G., 1995. Rheology of the Earth. Chapman and Hall, New York.\\
Richardson, D.C., Quinn, T., Stadel, J., Lake, G., 2000. Direct large-scale N-body simulations of planetesimal dynamics. Icarus 143, 45–59.\\
Rubin M., Altwegg K., Balsiger H., et al., 2015a. Molecular nitrogen in comet 67P/Churyumov-Gerasimenko indicates a low formation temperature. Science 348, 232–235.\\
Rubin, M. and 7 colleagues, 2020.\ On the Origin and Evolution of the Material in 67P/Churyumov-Gerasimenko.\ Space Sci. Rev. 216, 102.\\
Rugel, G. et al., 2009. New Measurement of the $^{60}$Fe Half-Life. Phys. Rev. Lett. 103, 072502.\\ 
Schmeling, H., Babeyko, A. Y., Enns, A., Faccenna, C., Funiciello, F., Gerya, T., Golabek, G. J., Grigull, S., Kaus, B. J. P., Morra, G., Schmalholz, S. M., van Hunen, J., 2008. A benchmark comparison of spontaneous subduction models–towards a free surface. Phys. Earth Planet. Inter. 171, 198–223.\\
Schubert, G., Spohn, T., Reynolds, R.T., 1986. Thermal histories, compositions and internal structures of the Moons of the Solar System. In: Satellites, J. A. Burns, M. S. Matthews (Eds.), 224–292, Univ. Arizona Press, Tucson.\\
Schwartz, S. R., Michel, P., Jutzi, M., Marchi, S., Zhang, Y., Richardson, D. C., 2018. Catastrophic disruptions as the origin of bilobate comets. Nat. Astron. 2, 379–382.\\
Sierks, H. et al., 2015. On the nucleus structure and activity of comet 67P/
Churyumov–Gerasimenko. Science 347, aaa1044.\\
Spencer, J. R., et al., The geology and geophysics of Kuiper Belt object
(486958) Arrokoth. Science 367, eaay3999.\\
Steckloff, J.~K., Lisse, C.~M., Safrit, T.~K., Bosh, A.~S., Lyra, W., Sarid, G., 2021.\ The Sublimative Evolution of (486958) Arrokoth. Icarus 356, 113998.\\
Tachibana, S., Huss, G. R., 2003. The initial abundance of $^{60}$Fe in the solar system. Astrophys. J. 588, L4144.\\
Tackley, P. J., Schubert, G., Glatzmaier, G. A., Schenk, P., Ratcliff, J. T., 2001. Three-dimensional simulations of mantle convection in Io. Icarus 149, 79–93.\\
Tillotson, J. H., 1962. Metallic equations of state for hypervelocity impact. GA-3216, General Atomic, San Diego, CA.\\
Vilenius, E. et al., 2014. “TNOs are Cool”: A survey of the trans-Neptunian region X. Analysis of classical Kuiper belt objects from Herschel and Spitzer observations. Astron. Astrophys. 564, A35.\\
Wahlberg Jansson, K., Johansen, A., 2014. Formation of pebble-pile planetesimals. Astron. Astrophys. 570, A47.\\
Wahlberg Jansson, K., Johansen, A., 2017. Radially resolved simulations of collapsing pebble clouds in protoplanetary discs. Mon. Not. R. Astron. Soc. 469, S149.\\
Wahlberg Jansson, K., Johansen, A., Bukhari Syed, M., Blum, J., 2017. The Role of Pebble Fragmentation in Planetesimal Formation. II. Numerical Simulations. Astrophys. J. 835, 109.\\
Weissman, P., Morbidelli, A., Davidsson, B., Blum J., 2020. Origin and Evolution of Cometary Nuclei. Space Sci. Rev. 216, 6.\\
Womack, M., Sarid, G., Wierzchos, K., 2017. CO and Other Volatiles in Distantly Active Comets. Publ. Astron. Soc. Pac. 129, 031001.\\
Womack, M., Curtis, O., Rabson, D. A., Harrington Pinto, O., Wierzchos, K., Cruz Gonzalez, S., Sarid, G., Mentzer, C., Lastra, N., Pichette, N., Ruffini, N., Cox, T., Rivera, I., Micciche, A., Jackson, C., Homich, A., Rosslyn Escoto, S., Erdahl, T., Goldschen-Ohm, M. P., Tollison, A., Reed, S., Zilka, J., Henning, B., Spinar, M., Uhl, W.T., 2021. The visual lightcurve of comet C/1995 O1 (Hale-Bopp) from 1995-1999. Planet. Sci. J. 2, 17.\\
Yomogida, K., Matsui, T., 1984. Multiple parent bodies of ordinary chondrites. Earth Planet. Sci. Lett. 68, 34–42.\\
Zhao, Y., Rezac, L., Skorov, Y., Hu, S.~C., Samarasinha, N.~H., Li, J.-Y., 2021. Sublimation as an effective mechanism for flattened lobes of (486958) Arrokoth.\ Nat. Astron. 5, 139–144.\\

\bibliographystyle{elsarticle-harv}
\balance
\endgroup

\pagebreak


\begin{table*}
\caption{Parameters of the four specific cases considered for the analysis. The impact velocity is 3 km/s and the impact angle is 45$^\circ$ for all cases. $M_{lr}/M_{tot}$ is the mass fraction of the largest remnant.}    
\label{table:4runs}      
\centering             
\begin{tabular}{c c c c }        
\hline\hline            
 Regime & Target radius &  Projectile radius & $M_{lr}/M_{tot}$ \\
 \hline        
Catastrophic & 50 km & 13.4 km & 0.66 \\
Cratering & 50 km  & 2.88 km & 0.997 \\
Catastrophic & 20 km  & 3.6 km & 0.59 \\
Cratering & 20 km & 0.8 km & 0.997 \\

\hline                               
\end{tabular}
\end{table*}


\begin{figure}
\begin{center}
\includegraphics[width=14cm]{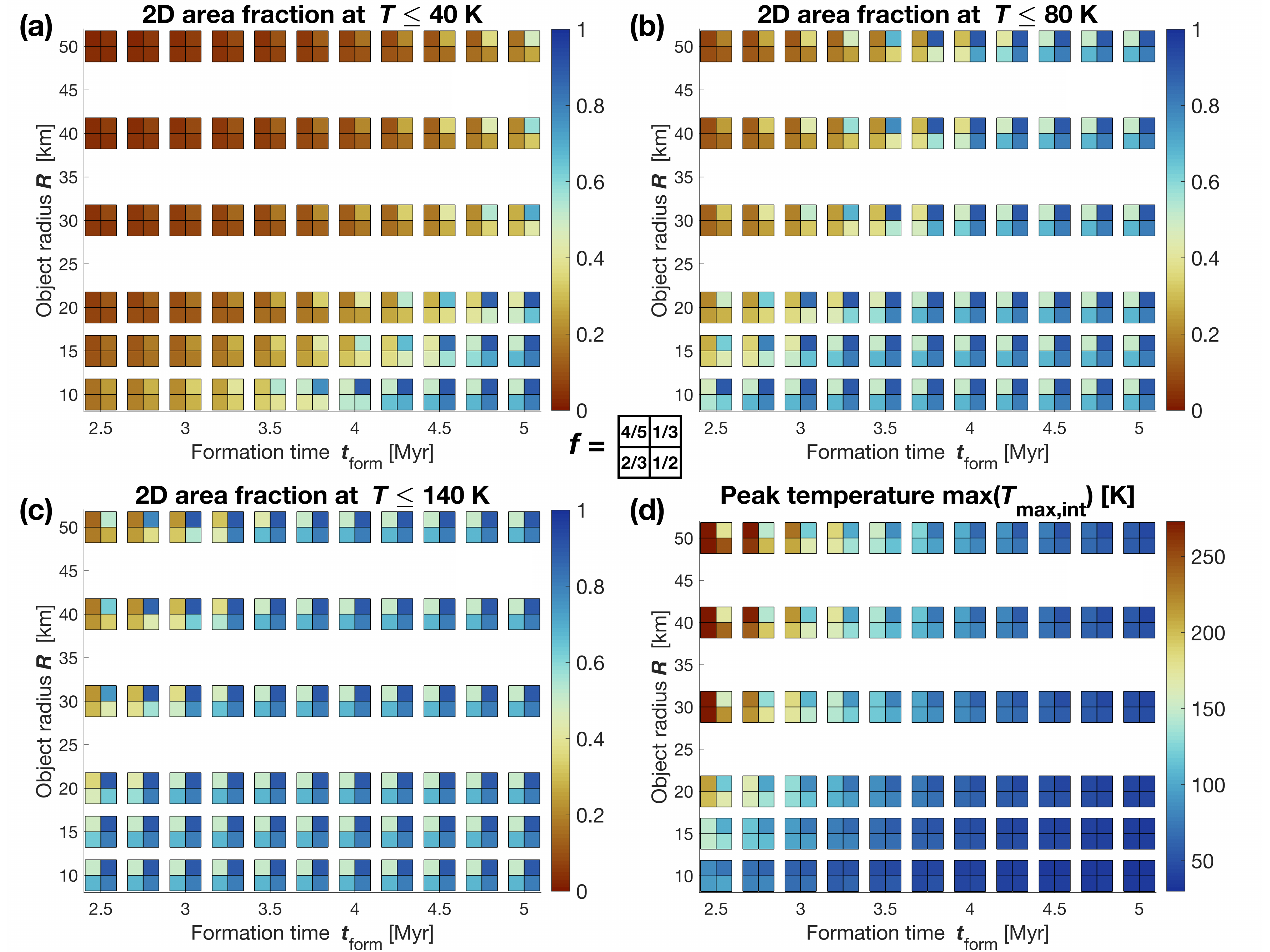}\\
\caption{2D area fraction inside icy planetesimal that experienced (a) $T \le 40$ K, (b) $T \le 80$ K and (c) $T \le 140$ K throughout the entire internal thermal evolution. Peak temperatures reached inside the icy planetesimal due to radiogenic heating are given in (d). Each of the boxes being part of a group of four in subplots (a) to (d) represents an icy planetesimal with same radius $R$ and formation time $t_{form}$, but different rock-to-ice mass fractions $f$ ranging clockwise from $1/3$ to $4/5$.}
\label{fig:}
\end{center}
\end{figure}

\begin{figure}
\begin{center}
\includegraphics[width=14cm]{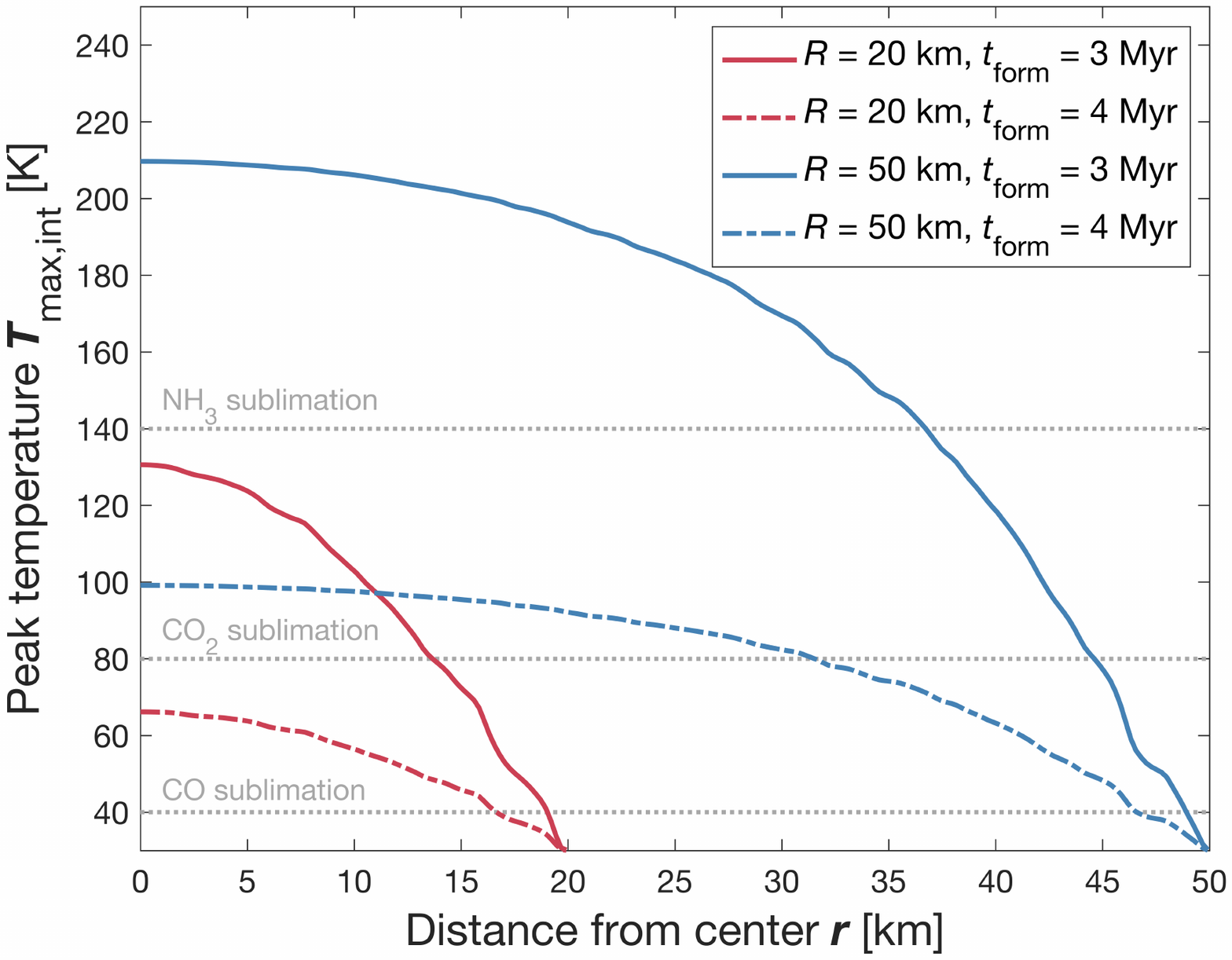}\\
\caption{Peak temperatures due to radiogenic heating at various distances from the center for icy planetesimals with 20 km (red) and 50 km radius (blue) and formation times $t_{form}$ of 3 Myr (solid lines) and 4 Myrs (dash-dotted lines) after CAI formation assuming a rock-to-ice mass fraction $f = 2/3$. Dotted gray lines denote the sublimation temperatures of CO, CO$_2$ and NH$_3$.}
\label{fig:peaktemp}
\end{center}
\end{figure}

\begin{figure}
\begin{center}
\includegraphics[width=14cm]{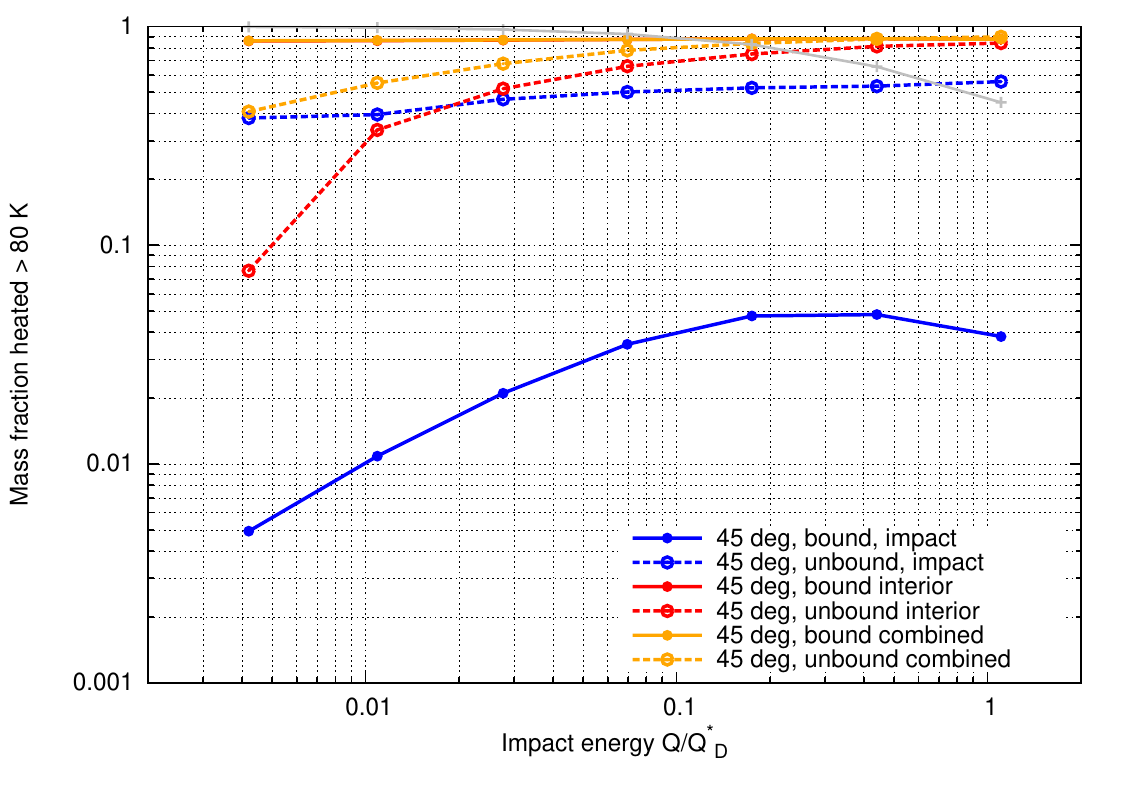}\\
\caption{Mass fraction of material that experienced $T_{max} > T_{crit}$ = 80 K by either impact or interior heating, as a function of normalized specific impact energy. Shown is the case with $R$ = 50 km, impact parameters $v_{imp}$ = 3 km/s, $\theta$ = 45$^\circ$ and a formation time of $t_{form}$ = 3 Myr. The gray line shows the relative size of the largest remnant. We note that the solid orange line lies on top of the solid red line.}
\label{fig:heating_vs_q_t0_3ma}
\end{center}
\end{figure}

\begin{figure}
\begin{center}
\includegraphics[width=14cm]{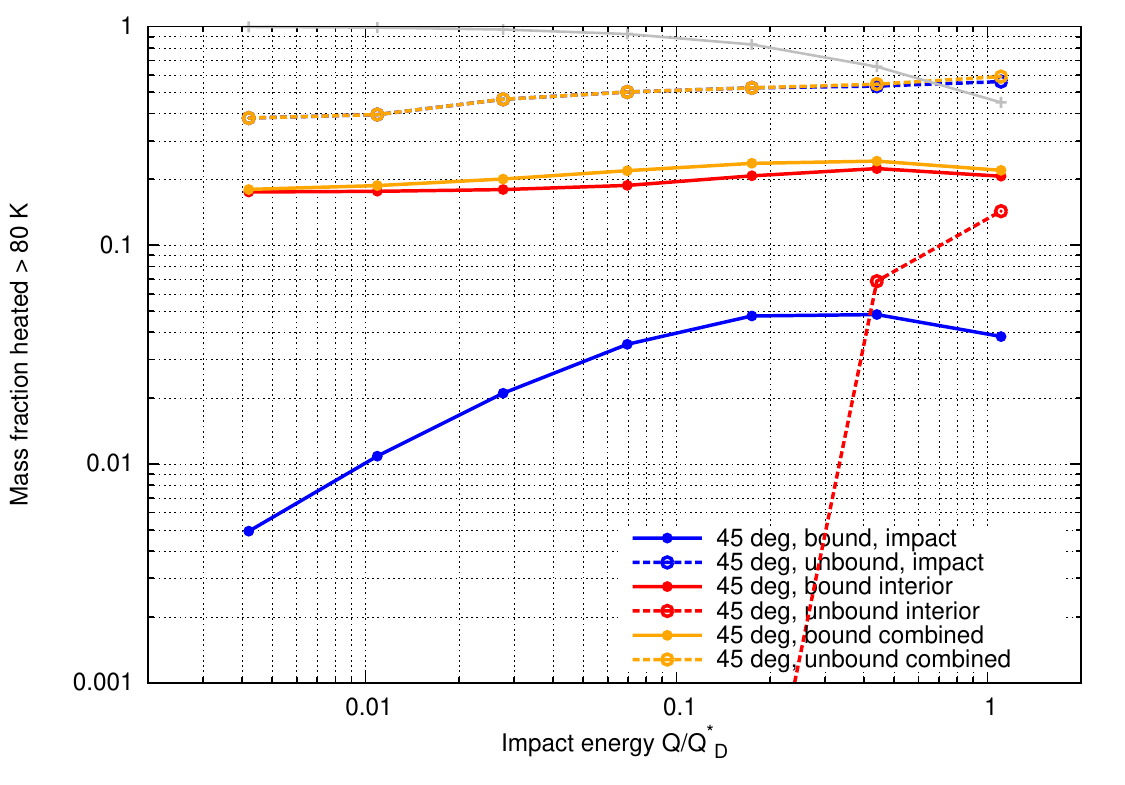}\\
\caption{Mass fraction of material that experienced $T_{max} > T_{crit}$ = 80 K by either impact or interior heating, as a function of normalized specific impact energy. Shown is the case with $R$ = 50 km, impact parameters $v_{imp}$ = 3 km/s, $\theta$ = 45$^\circ$ and a formation time of $t_{form}$ = 4 Myr. The gray line shows the relative size of the largest remnant.}
\label{fig:heating_vs_q_t0_4ma}
\end{center}
\end{figure}

\begin{figure}
\begin{center}
\includegraphics[width=14cm]{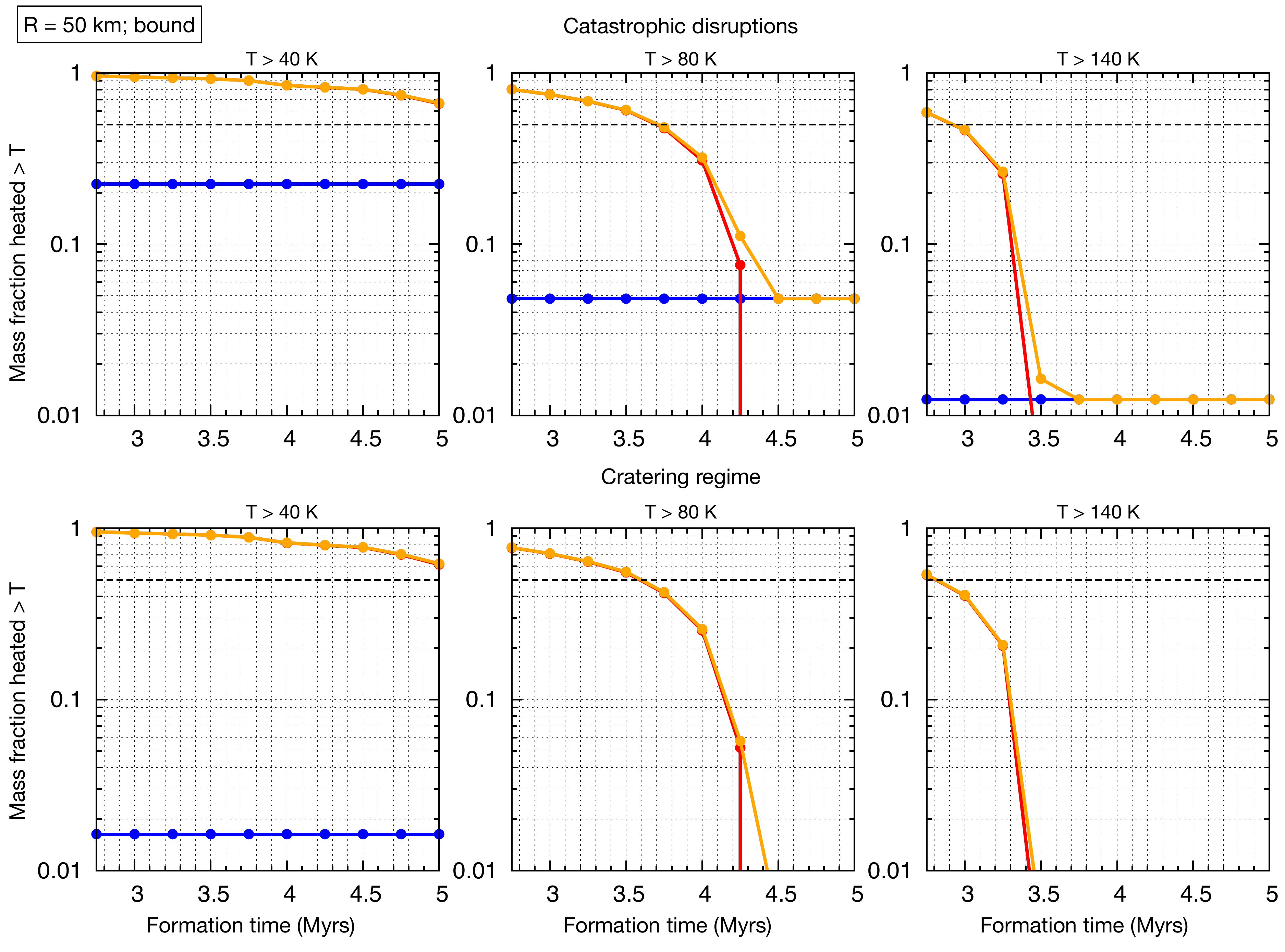}\\
\caption{Mass fraction heated to $T>$ 40 K (left column), $T >$ 80 K (middle column) and $T>$ 140 K (right column) vs. formation time $t_{form}$ considering bound material originating from $R=$ 50 km icy planetesimals. Catastrophic disruptions (top panel) and cratering events (bottom panel). Red: Material heated by interior evolution, blue: Material heated by impact event, orange: Material heated by either internal evolution or by impact. The dashed horizontal line indicates a mass fraction of 0.5. We note that in the $T>$ 40 K plots, the orange line lies on top of the red line.}
\label{fig:t_r_50_bound}
\end{center}
\end{figure}

\begin{figure}
\begin{center}
\includegraphics[width=14cm]{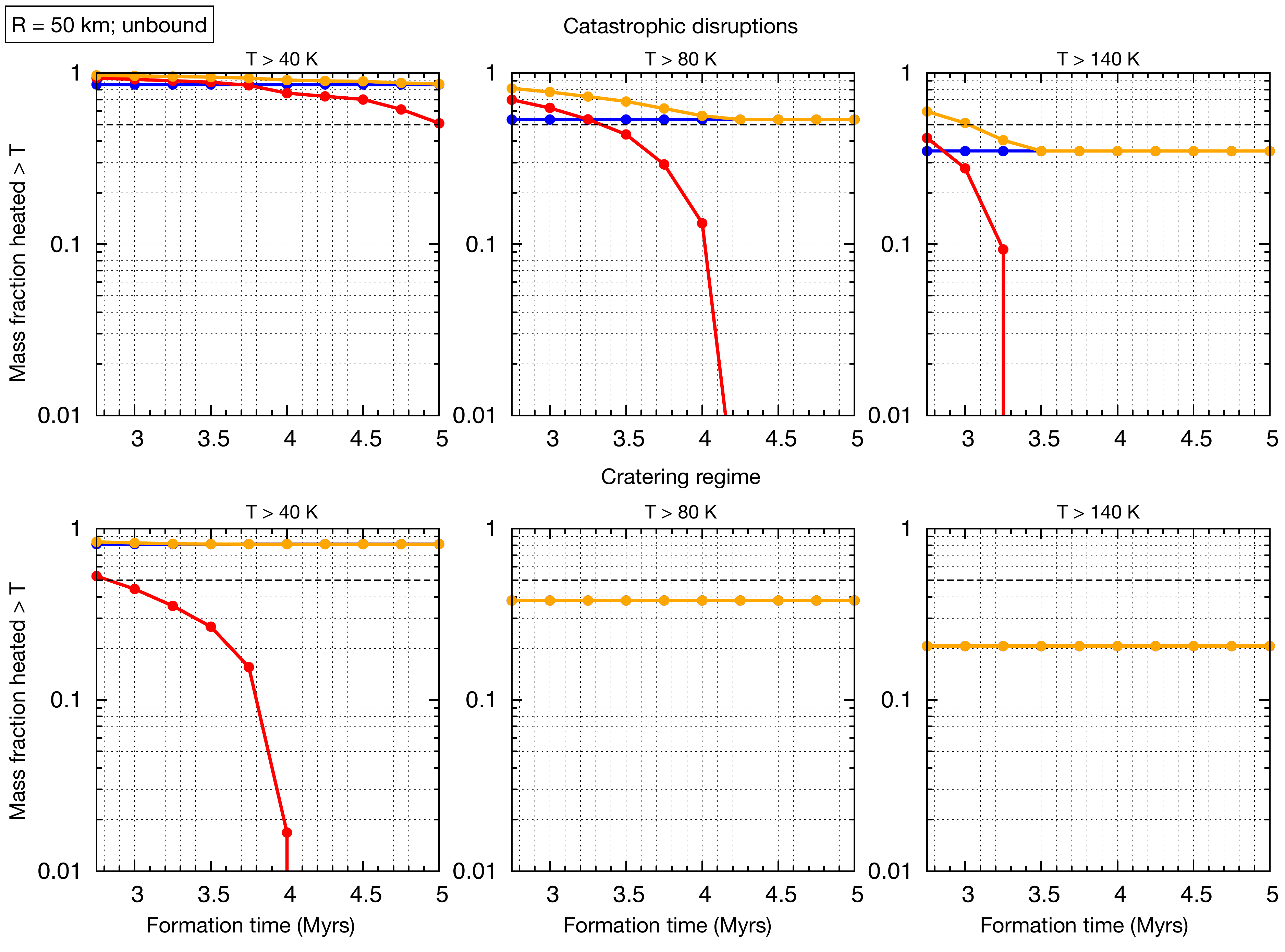}\\
\caption{Mass fraction heated to $T>$ 40 K (left column), $T >$ 80 K (middle column) and $T>$ 140 K (right column) vs. formation time $t_{form}$ considering unbound material originating from $R=$ 50 km icy planetesimals. Catastrophic disruptions (top panel) and cratering events (bottom panel). Red: Material heated by interior evolution, blue: Material heated by impact event, orange: Material heated by either internal evolution or by impact. The dashed horizontal line indicates a mass fraction of 0.5. We note that in the cratering regime plots, the orange line lies on top of the blue line.}
\label{fig:t_r_50_unbound}
\end{center}
\end{figure}

\begin{figure}
\begin{center}
\includegraphics[width=14cm]{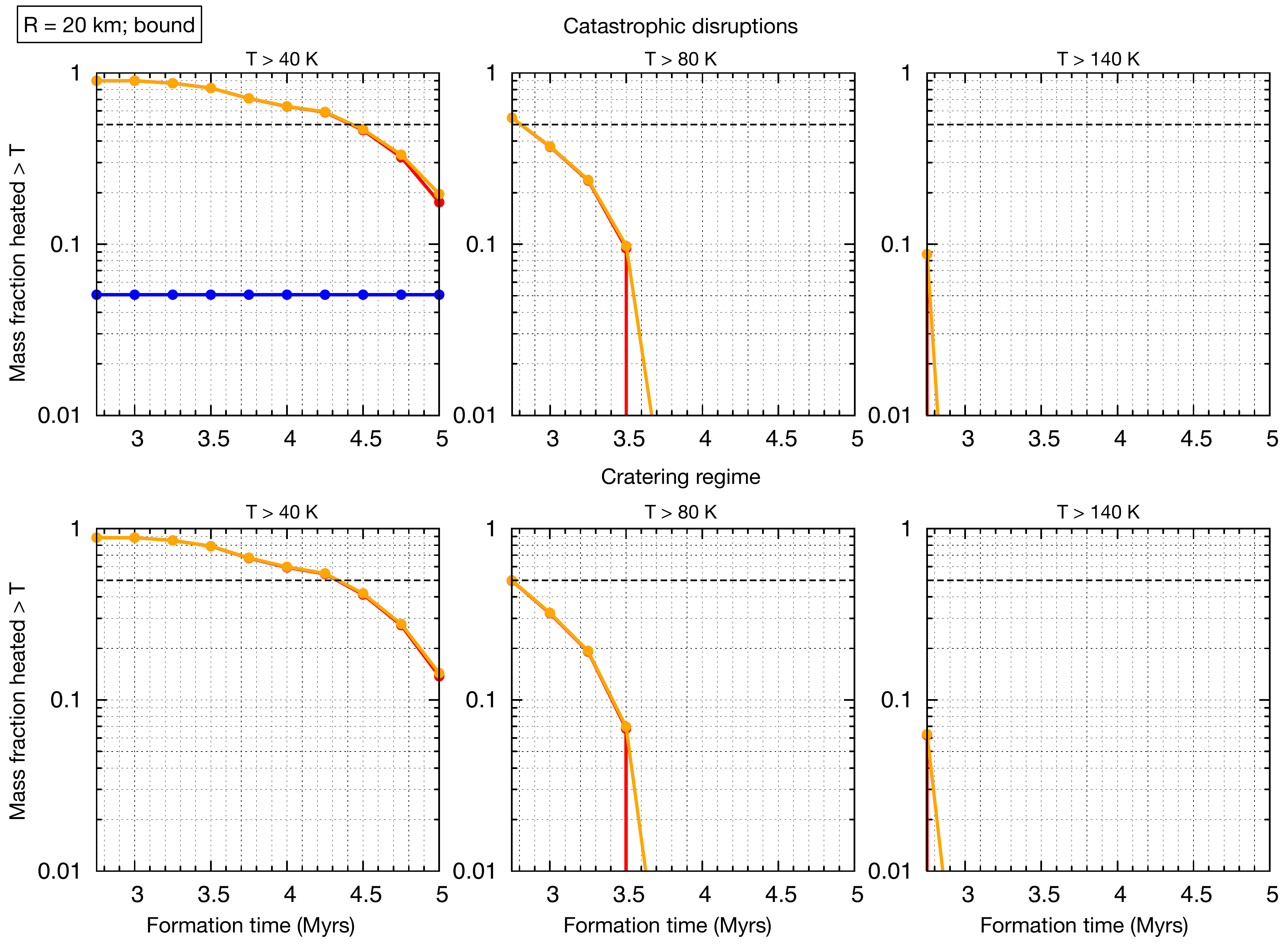}\\
\caption{Mass fraction heated to $T>$ 40 K (left column), $T >$ 80 K (middle column) and $T>$ 140 K (right column) vs. formation time $t_{form}$ considering bound material originating from $R=$ 20 km icy planetesimals. Catastrophic disruptions (top panel) and cratering events (bottom panel). Red: Material heated by interior evolution, blue: Material heated by impact event, orange: Material heated by either internal evolution or by impact. The dashed horizontal line indicates a mass fraction of 0.5.}
\label{fig:t_r_20_bound}
\end{center}
\end{figure}

\begin{figure}
\begin{center}
\includegraphics[width=14cm]{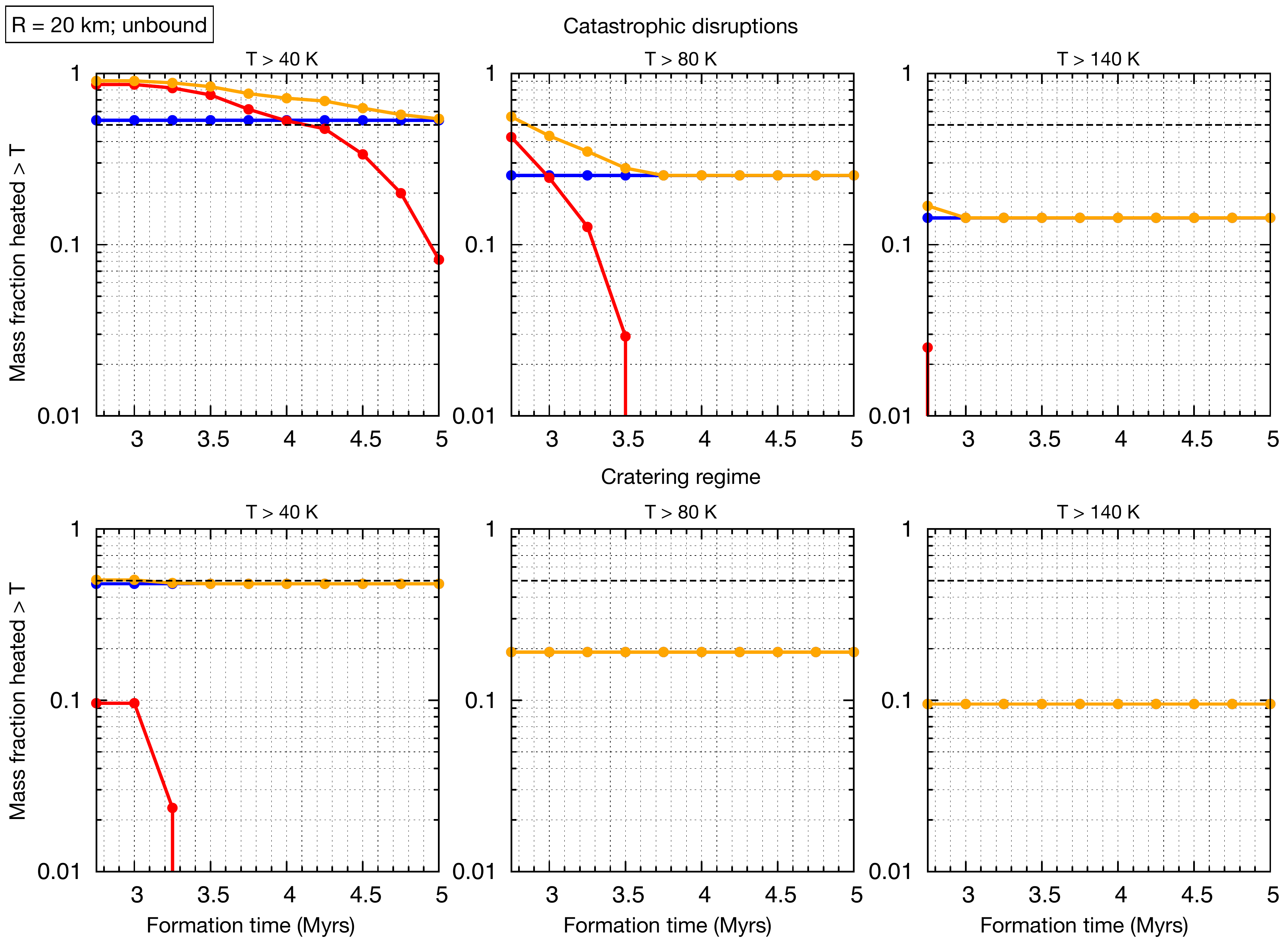}\\
\caption{Mass fraction heated to $T>$ 40 K (left column), $T >$ 80 K (middle column) and $T>$ 140 K (right column) vs. formation time $t_{form}$ considering unbound material originating from $R=$ 20 km icy planetesimals. Red: Material heated by interior evolution, blue: Material heated by impact event, orange: Material heated by either internal evolution or by impact. The dashed horizontal line indicates a mass fraction of 0.5. We note that in the cratering regime plots, the orange line lies on top of the blue line.}
\label{fig:t_r_20_unbound}
\end{center}
\end{figure}

\begin{figure}
\begin{center}
\includegraphics[width=14cm]{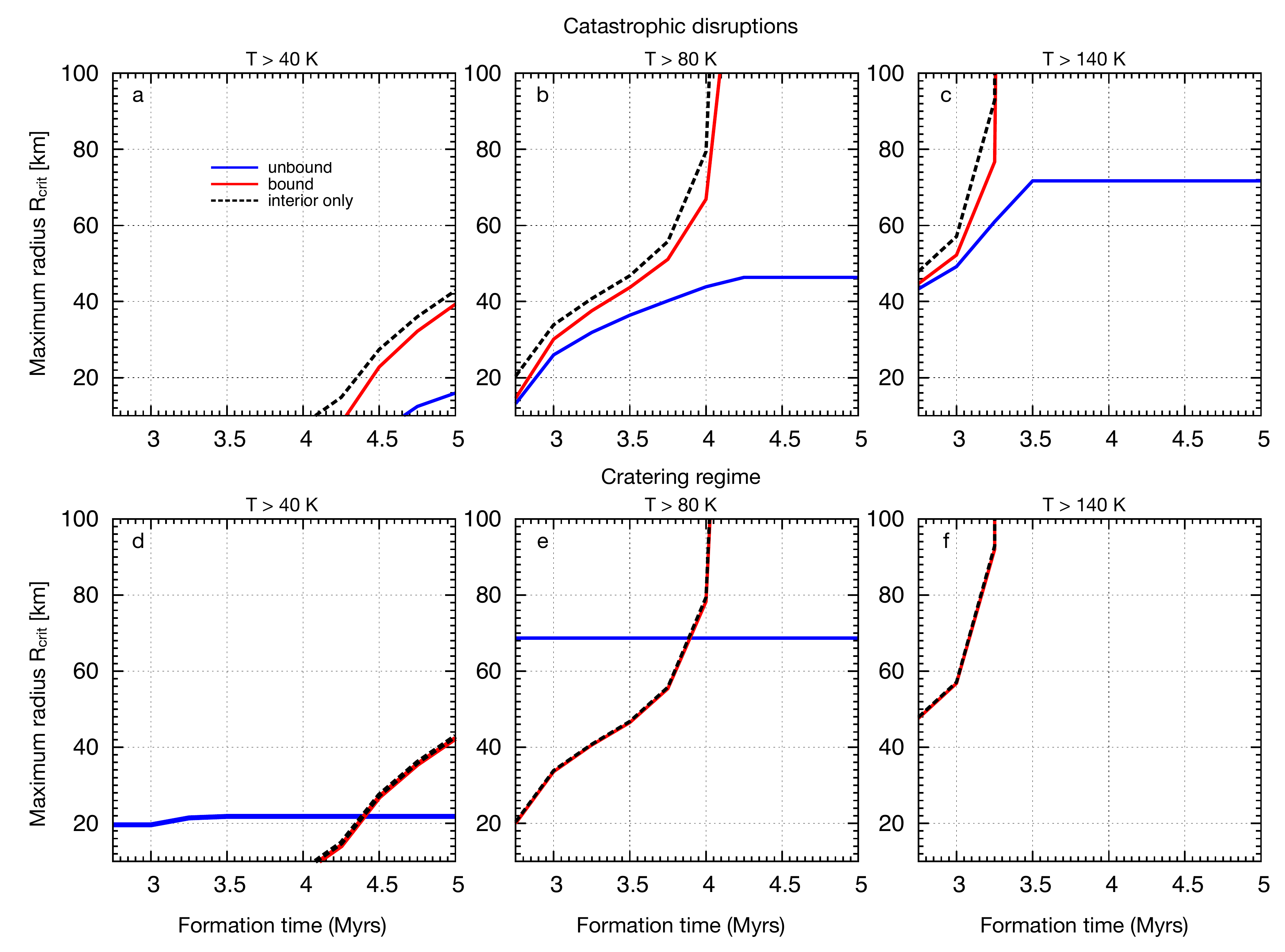}\\
\caption{Formation time vs. potential maximum radius of icy planetesimal for $f_{heated}$ = 0.5 assuming critical temperatures of $T>$ 40 K (left column), $T>$ 80 K (middle column) and $T>$ 140 K (right column) for catastrophic disruptions (top panel) and cratering events (bottom panel). Blue: Comet/KBO originates from unbound post-impact material, red: Comet/KBO originates from bound post-impact material, black: Comet/KBO corresponds to primordial icy planetesimal.}
\label{fig:rcrit_tform}
\end{center}
\end{figure}

\begin{figure}
\begin{center}
\includegraphics[width=14cm]{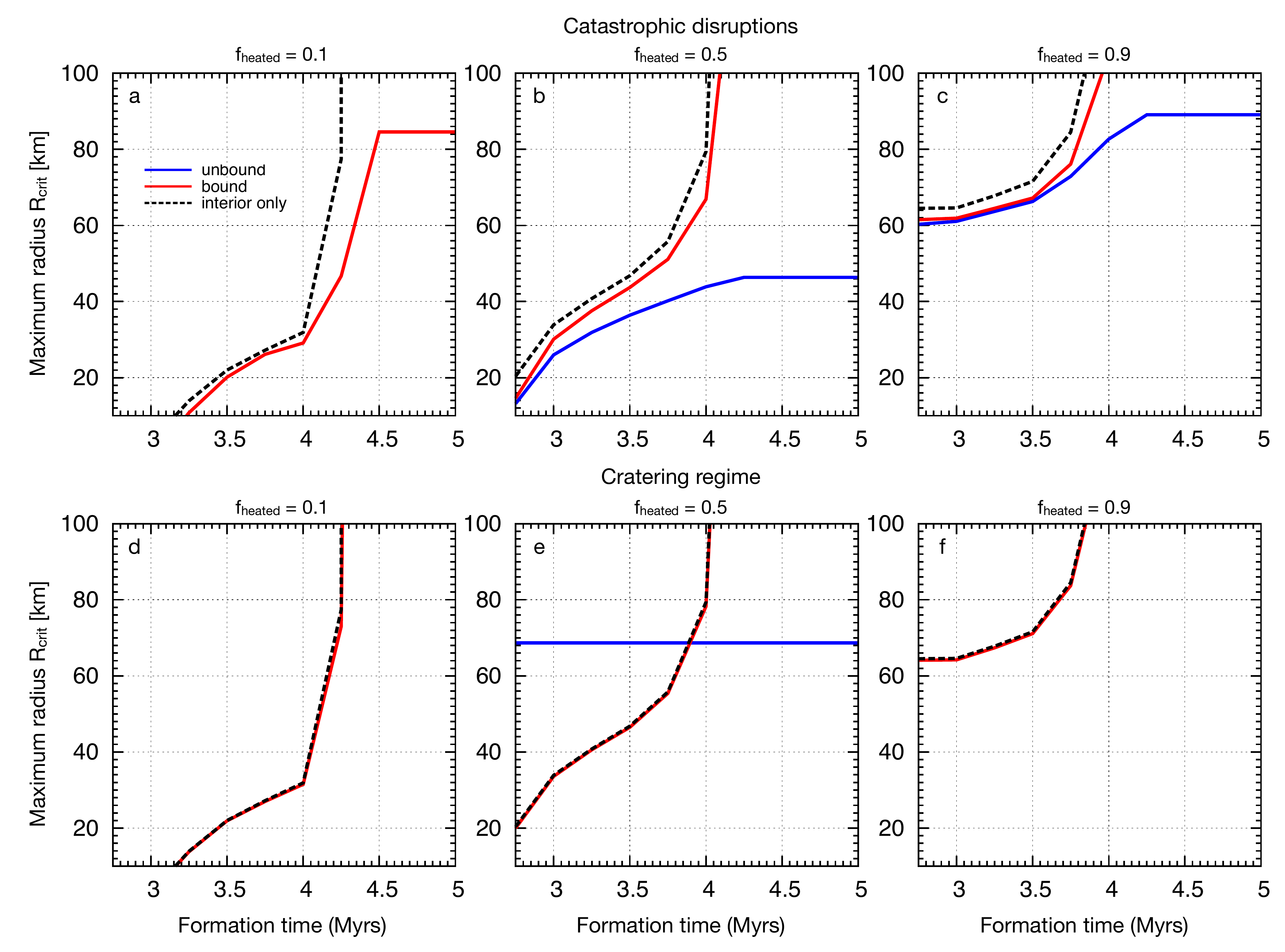}\\
\caption{Formation time vs. potential maximum radius of icy planetesimal for $f_{heated}$ = 0.1 (left column), $f_{heated}$ = 0.5 (middle column) and $f_{heated}$ = 0.9 (right column) assuming a critical temperatures of $T>$ 80 K for catastrophic disruptions (top panel) and cratering events (bottom panel). The nominal case of $f_{heated}$ = 0.5 corresponds to Figure \ref{fig:rcrit_tform}b+e. Blue: Comet/KBO originates from unbound post-impact material, red: Comet/KBO originates from bound post-impact material, black: Comet/KBO corresponds to primordial icy planetesimal.}
\label{fig:rcrit_tform_fractions}
\end{center}
\end{figure}

\end{document}